# MOTION-BASED GENERATORS FOR INDUSTRIAL APPLICATIONS

*T. Sterken[1,2], P. Fiorini[1], R. Puers[2]*

[1] IMEC, Kapeldreef 75, 3001 Leuven, Belgium
[2] ESAT, K.U.Leuven, Kasteelpark Arenberg 10, 3001 Leuven, Belgium

## ABSTRACT

*Scaling down of electronic systems has generated a large interest in the research on miniature energy sources. In this paper a closer look is given to the use of vibration based scavengers in industrial environments, where waste energy is abundantly available as engine related vibrations or large amplitude motions. The modeling of mechanical generators resulted in the design and realization of two prototypes, based on electromagnetic and electrostatic conversion of energy. Although the prototypes are not yet optimized against size and efficiency, a power of 0.3 mW has been generated in a 5 Hz motion with a 0.5 meter amplitude.*

## INTRODUCTION

Since the presentation of the first transistor at the Bell Laboratories the size of electronic components has decreased according to Moore's law, resulting in highly functional computers on miniature size silicon dies today. At the same time the integration of systems on chips (SOCs) and the development of systems in packages (SIPs) decrease the overall size of electronic devices, transforming them into small, lightweight and highly functional systems. This miniaturization has triggered new desires: mobility, autonomy and wireless connectivity are key features to be taken into account when designing a new system. These systems also require a wireless, autonomous and mobile energy source. The conventional solution is found in storage systems such as electrochemical batteries. The total amount of energy available depends on the energy density and the volume of the storage system. New storage systems have been developed based on higher energy densities in order to decrease the volume and weight occupied by the energy source (e.g. fuel cells). The volume-dependency of the amount of stored energy is however an inevitable drawback. As a consequence, when system dimension decrease, the smaller volume available for energy storage results in a shortened lifetime. A solution to this contradiction is offered by miniature generators, extracting energy from the environment of the mobile system and converting it efficiently into electrical power. Several literature contributions present optimistic evaluations of the performances of this type of *waste energy generators*, based on solar, thermal or mechanical energy [1]. Industrial environments require small lightweight sensor devices to monitor the processes and the condition of the machinery. In order to prevent the costs of battery replacements, waste energy generators provide a good solution. An ideal source of energy for these *scavengers* is offered by the mechanical energy available from the "natural" motion and from vibrations of the machinery. The first category of mechanical energy is the result of robotic movements and is characterized by low frequencies (up to 15 Hz or 900 rpm) and large amplitudes of the order of 1 meter. Machinery vibrations have harmonic frequencies up to 5 kHz but exhibit much smaller amplitudes.

## MODELING MECHANICAL ENERGY SCAVENGERS

In order to extract mechanical energy from a motion, it is necessary to convert this motion into a relative movement across the mechanical port of the generator. This is not a trivial task when the dimensions of the generator (and of its parts) become small. Miniature generators therefore use the inertia of a mass to convert the source motion into a local displacement across the generator. The bigger the mass, the better the motion is coupled onto the generator; yet again the need for miniaturization limits the size of the mass. The motion of the mass is damped by the generator, while it is extracting energy from the motion, but also by unwanted factors such as air damping, squeezed film damping and friction, which convert the kinetic energy into heat. The inertial mass is often suspended to reduce these losses and to improve the efficiency of the device.

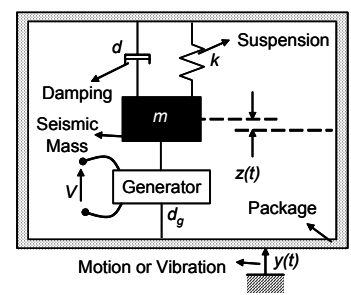

*Fig. 1. Overview scheme for mechanical energy scavengers*

The scheme in Figure 1 is completed with the addition of a spring to model the spring-like behavior of the suspension. The combination of a mass suspended by springs allows the designer to tune the device into resonance at a desired frequency, e.g. the one of the vibration source.

The mechanical lumped element model in Figure 1 allows an initial estimation of the generated power as a function of the input parameters $Y$ (amplitude of the vibration) and $\omega$ (rotational frequency of the vibration). We assume that both the generator and the unwanted damping counteract the motion of the mass with a force linearly proportional to the velocity $\dot{Y}$ of the mass. The power is then given by:

$$Power = \frac{d_g m^2 \omega^4}{(k - m\omega^2)^2 + 4\omega^2(d + d_g)^2} \frac{|\dot{Y}|^2}{2} \qquad (1)$$

In this equation the average power is a function of the mass $m$, the spring constant of the suspension $k$ and the damping coefficients $d$ and $d_g$ of the unwanted and the generator based damping respectively. Note that the vibration is assumed to be sinusoidal, although the calculation could be repeated for any type of input motion. Expression (1) can be rewritten as:

$$Power = m\omega \frac{|\dot{Y}|^2}{2} \frac{\zeta_g x^3}{(1 - x^2)^2 + x^2(\zeta + \zeta_g)^2} \qquad (2)$$

considering the following substitutions:

$$\omega_{res} = \sqrt{k/m} \qquad x = \omega/\omega_{res} \qquad \zeta_{(g)} = \frac{d_{(g)}}{2\sqrt{km}}$$

The expression (2) becomes dimensionless if it is divided by $\frac{1}{2}m\omega|\dot{Y}|^2$: the generated power depends linearly on the weight (and thus the size) of the device.

### 2.1.1. Generator modeling

In the lumped mechanical model of figure 1 the generator has been modeled as a mechanical linear damper, although the energy is converted into electrical energy. It is useful to develop a model that allows the study of the energy-flow from the mechanical to the electrical domain and vice-versa, to investigate the effects of different electrical load circuits on the mechanical behavior, and thus on the generated electrical energy.

In this study we choose to translate the mechanical lumped model towards an electrical equivalent circuit,

where currents *(i)* and voltages *(v)* represent velocities *(ż)* and forces *(F)* respectively. The equivalence translates masses into inductances, while springs and dampers are represented by capacitors and resistors.

An equivalent electrical model allows the use of circuit analysis software such as SPICE to analyze the mechanical behavior of the system as well as the electrical influences on the behavior. The equivalent mechanical model is coupled to the load circuit by an equivalent model of the generator, that consists of a transformer between the mechanical parameters *(F,ż)* and the electrical parameters *(v,i)*, in parallel with a capacitor [2]. This capacitor in parallel is replaced by an inductance in series when an electromagnetic generator is used (Figure 2).

Note that both the models for electrostatic, piezo-electric and electromagnetic generators are linear models: non-linear relations between *(F,ż)* and *(v,i)* were linearized, e.g. when modeling electrostatic generators [2].

### INDUSTRIAL MOTION AND VIBRATIONS

Mechanical waste energy in industrial environments originates from a variety of situations, ranging from vibrations of rotating engines to the movement of machine parts. The source of energy that will be used depends strongly on the application, which might be a very small and lightweight sensor, with low power but high mobility requirements, or could be a stationary sensor inside the machine.

Table 1 gives a short overview of some example vibrations in industrial environments with different frequencies and amplitudes/accelerations. The last column of the table indicates the value $\frac{1}{2}\omega_{input}|\dot{Y}|^2$.

The generated power is obtained from this value by multiplying it with the mass of the generator and the dimensionless term of expression (2), again considering a tuning of the resonance frequency of the system to the input frequency $\omega_{input}$. The last column is therefore a good measure to compare the energy available in different motions and vibration, independently of the actual design of the generator that is used.

The values in Table 1 indicate the wide range of available power in industrial environments, and the necessity to obtain a high mass. The contradiction between power and size is once more illustrated.

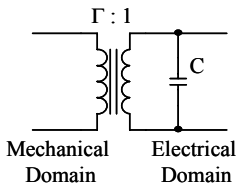

Fig. 2a. Equivalent model for electrostatic and piezo-electric transducers.

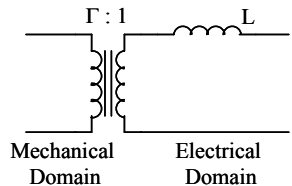

Fig. 2b. Equivalent model for electromagnetic transducers.

| Source Type | Amplitude / Acceleration | Frequency | $\frac{1}{2}\omega_{input}|\dot{Y}|^2$ |
|---|---|---|---|
| Machine Vibration | 1 m/s² | 1000 Hz | 79.5 µW/kg |
| | 10 m/s² | 200 Hz | 39.8 mW/kg |
| Machine Motion | 0.25 m | 2 Hz | 62 W/kg |
| | 0.5 m | 5 Hz | 3.87 kW/kg |

Table 1: Typical examples of industrial vibration/motion

*T. Sterken, P. Fiorini, R. Puers*

# MOTION-BASED GENERATORS FOR INDUSTRIAL APPLICATIONS

### 3.1. Small Amplitude Vibrations

The first two examples of Table 1 refer to vibrations on and in the neighborhood of machinery. In both cases the reference power $\frac{1}{2}\omega_{input}|\hat{Y}|^2$ is low. As masses are always limited by the dimensions, the designer needs to use the resonance in the dimensionless term of expression (2) to achieve enough energy to run the application. Although this strategy is often reported in literature, it is again limited by the restricted size of the scavenger: the use of resonance results in a large internal displacement of the mass, and the motion is limited by the impact of the mass onto the package of the device.

One way to prevent this impact is to damp the generator electromechanically by applying a correct load. The optimum power is then given by [3]:

$$Power = m\omega_{input} \frac{|\hat{Y}|^2}{2} \frac{Z_{max}}{Y} \qquad (2)$$

The maximum allowed amplitude of the motion of the mass is given by $z_{max}$. In some cases it is however not possible to damp the movement of the mass by applying the optimal load, as fabrication restrictions prevent a good electromechanical coupling, and thus electromechanical damping.

Another strategy that is applied in those cases is to allow impact, resulting in a higher power, often delivered at better voltage levels (or current levels, depending on the type of generator), but at the cost of reduced reliability [4].

In Figure 3 a design for an electrostatic generator intended for small amplitude vibrations is shown. It consists of a variable overlap capacitor positioned between the movable mass and the bottom wafer. The mass is suspended by springs (not shown in Figure 3).

The capacitor is polarized by bonding an *electret* on top of the capacitor. This is a dielectric material that has been charged, in order to create a fixed and stable electric field across the capacitor. A capacitance change of this polarized capacitor will thus result in a current through the load that is applied between the electret and the fixed electrode.

A variable overlap capacitor of this type has promising values for electromechanical coupling, especially when the device is scaled to small dimensions [5], however the fabrication of the device requires advanced techniques in wafer bonding and bulk micromachining: a small capacitor gap needs to be ensured between the silicon wafer and the glass bottom wafer. This gap was realized by bonding the wafers with photosensitive BCB, which allows the reduction of the gap down to 1 µm. This bond has the advantage of being resistant to the KOH-based etch-step that is used to bulk micromachine the mass of the generator. The suspensions of the mass need to be stiff in the direction perpendicular to the motion in order to prevent pull-in of the mass during operation. However, the same suspension must be flexible in the direction of the vibration to allow the movement of the mass at the correct frequency. This contradiction is resolved by using deep reactive ion etching with an aspect ratio up to 20.

In figure 4 a SEM picture of the released generator is shown. The mass has been released, and the capacitance has been measured as a function of the displacement of the mass. The graph of figure 5 shows the measured capacitance as well as the simulated capacitance. The measured capacitance has a slightly higher peak value as well as a higher stray field capacitance. This might indicate the presence of a parasitic capacitance. The curve illustrates a capacitance change of 0.09 pF/µm where a maximum value of 0.23 pF/µm was expected. These values can be improved by optimizing the design towards surface: the current capacitances occupy merely 0.6 mm². Enlargement of the surface scales both capacitances and capacitance changes.

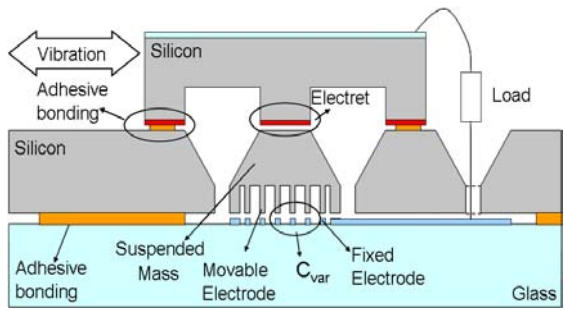

*Fig. 3. Schematic of an electrostatic generator for small amplitude movements*

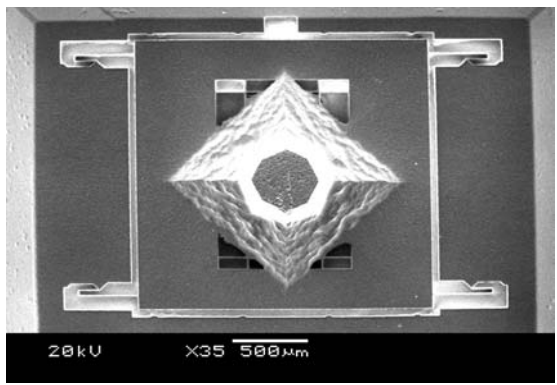

*Fig. 4. SEM photograph of the processed electrostatic generator*

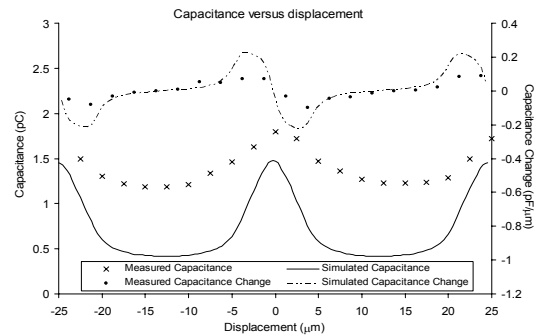

*Fig. 5. Capacitance and capacitance change per micrometer*

*T. Sterken, P. Fiorini, R. Puers*

# MOTION-BASED GENERATORS FOR INDUSTRIAL APPLICATIONS

### 3.2. Large Amplitude Vibrations

An electromagnetic generator was chosen to prototype a generator for large motion applications in industry. It consists of a wired tube, in which a rotor of magnets is able to move linearly. The coil consists of a series-connection of smaller coils, whose lengths equal the length of the rotor magnet. At every connection between two coils the winding direction is changed (Figure 6a). This winding strategy ensures that the voltage generated by the flux-change of one pole of the magnet does not counteract the voltage generated by the other pole of the magnet, moving at the same speed, yet through a alternately wound coil. No spring is used: resonance does not improve the behavior of the generator when the input amplitude is larger than the dimensions of the generator.

Two prototypes where build, based on 1T NdFeB cylindrical rotors with diameters of 4 mm and 10 mm.

A wire with a diameter of 100 μm (AWG 39) was used, changing direction every 3 mm and 16 mm respectively. The SPICE model for the electrostatic generator was adapted for modeling the effects of impact [4] and the influence of alternating wiring directions. The simulation of a 2 Hz motion at an amplitude of 25 cm, and the corresponding measurement are given in Figure 7.

The second prototype was mounted on an industrial tool that moves the generator over 0.5 meter at a speed of 5 Hz. The load of the generator consisted of a voltage doubling circuit, which charges 2 capacitors of 220 μF.

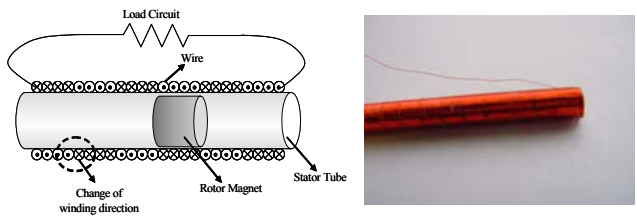

*Fig. 6a. Wiring scheme and prototype realization (Fig. 6b) of a linear electromagnetic generator*

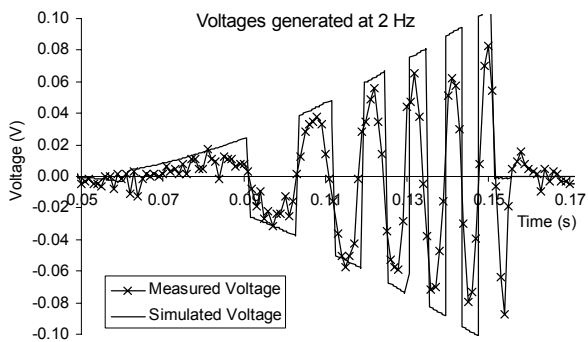

*Fig. 7. Simulation and measurement of a 25 cm motion at 2Hz.*

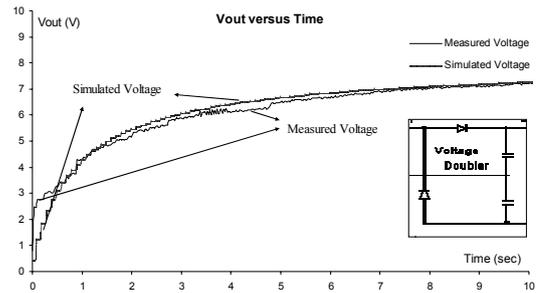

*Fig. 8. Simulation and measurement of a 1 meter motion at 5Hz, charging a capacitor through a voltage doubling circuit (inset).*

The curve in Figure 8 illustrates the charging over time of the capacitances during the simulation as well as during the measurement. The voltage is measured across both capacitances at once. The energy stored in the capacitances after 10 seconds is 2.9 mJ, indicating a generated power level of 290 μW.

### 5. CONCLUSIONS

Motion and vibration in industrial environments are reliable energy sources for mechanical waste energy generators. Depending on the type of motion different generators have been designed and fabricated, based on electrostatic and electromagnetic generators. Initial tests of these prototypes on industrial machinery illustrate the coherence between modeling and measurements.

### 6. ACKNOWLEDGEMENTS

Research funded by a Ph.D grant of the Institute for the Promotion of Innovation through Science and Technology in Flanders (IWT-Vlaanderen).

The authors would like to thank Bram Cuvelier, Aissa Couvreur, Steven Sanders and Lieven Hollevoet for their help and guidance in testing the prototypes.